\documentclass[aps,showpacs]{revtex4}
\usepackage{amsmath}
\usepackage{epsfig}

\begin{document}
\title{Application of the Lorentz-Transform Technique to Meson Photoproduction}

\author{Christoph Rei{\ss}$^{1,2,3}$,
  Winfried Leidemann$^{1}$, 
  Giuseppina Orlandini$^{1}$ 
  and Edward L. Tomusiak$^{2}$}
\affiliation{$^{1}$Dipartimento di Fisica, Universit\`a di Trento, and
  Istituto Nazionale di Fisica Nucleare, Gruppo Collegato di Trento,
  I-38050 Povo, Italy\\
  $^{2}$Department of Physics and Astronomy
  University of Victoria, Victoria, BC V8P 1A1, Canada\\
  $^3$present address: Institut f\"ur Kernphysik, Universit\"at Mainz,
  Johann-Joachim-Becher-Weg 45, D-55099 Mainz, Germany\\
}

\date{\today}

\begin{abstract}
  \noindent
  We show that the Lorentz integral transform ( LIT ) technique which has been
  successfully applied to photoreactions in light nuclei can also be
  applied to photoreactions involving particle production. A simple model where
  results are easily calculable in the traditional fashion is used to test
  the technique. Specifically we compute inclusive $\pi^+$ photoproduction from
  deuterium for photon energies less than 200 MeV using a Yamaguchi model for
  the NN interaction.  It is demonstrated that although the response functions
  for inclusive meson production do not have favourable asymptotic behavior
  one can nontheless extract them by inversion of the transform.  
  The implication is that one can treat realistic problems 
  of photo-meson production including all final state interactions 
  by means of the LIT technique.
\end{abstract}
\bigskip
\maketitle

\section{Introduction}

In a series of papers \cite{elo1,elo2,elo3,elo4,elo5,elo6,mkog} it has 
been demonstrated that the LIT technique allows a convenient calculation
of inclusive photoreaction cross sections wherein final state interactions
are fully included.  Further, the technique has been extended \cite{efros85,lapwl} to
exclusive reactions.  A merit of the technique is that the
calculation of continuum wave functions is avoided.  In fact the
differential equations to be solved are inhomogeneous and have
solutions bounded at infinity.  The work cited above is based on
non-relativistic quantum mechanics and a nucleons-only subspace.

In order to treat meson photoproduction from nuclei it is desirable to
see if the LIT technique can be extended to 
a larger subspace i.e. one including nucleons and mesons.  This would
enable the full inclusion of both meson-nucleon and NN interactions 
in the production calculations.  
As a test we consider inclusive photoproduction of low energy ( $<40$ MeV )
$\pi^+$ mesons from
deuterium.  A traditional calculation of this process was reported
by Dressler, MacDonald, and O'Connell\cite{Dress} and by Gupta,
Anand, and Bhasin \cite{Gupta}. Both of these groups employed a simple 
Yamaguchi potential \cite{Yam} with parameters chosen to account for
deuteron binding and low energy NN scattering properties.  We adopt this
model with the parameters used in \cite{Dress}. 
Thus this example attempts to apply the LIT technique
in the NN$\pi$ subspace where for simplicity only an NN interaction is included.
Although no $\pi-$N scattering is contained in this
model, this is not essential for testing the method. 
Our results will show that it is indeed possible to apply the LIT
method to this problem and to extract the total photoproduction
cross section. We should point out that our restriction to a simple
NN potential for this test calculation does not modify these conclusions.
This is clearly seen from the series of papers \cite{elo1,elo2,elo3,elo4,elo5,elo6,mkog}
where a range of NN potentials, from simple phenomenological models to
modern realistic potentials, have been used in 3-nucleon and 4-nucleon
problems.  In no case did the complexity of the NN potential model
have any effect on the accuracy or implementation of the LIT technique.
With assurance that the formalism works the addition
of a meson-nucleon interaction can be viewed as a technical point, the
consequence of which would be more complicated numerics.
 
Low energy
photoproduction is described by the Kroll-Ruderman \cite{KR} operator
\begin{eqnarray}
  {\cal H}_{\rm int}(x)
  =
  -i\, e\, \left(\frac{f}{m_\pi}\right)\,
  \sum_j\,\hat\epsilon_\lambda\cdot\vec\sigma(j)\,\tau_-(j)\,
  e^{-i\, k\cdot x}\,\phi_+(x)\,\delta(\vec x-\vec x_j)
  \quad.
\end{eqnarray}
where $\vec k$ and $\hat\epsilon_\lambda$ 
denote the incident photon momentum and polarization vectors
, respectively,
and $\phi_+(x)$ 
denotes 
the meson field operator
\begin{eqnarray}
  \phi_+(x)
  =
  \frac{1}{\sqrt{8 \pi^3}}
  \int\!{\rm d}^3q\,\,
  \frac{1}{\sqrt{2 q_0}}\,
  \left[
    e^{iq\cdot x}\,a_+^\dag(\vec q\,) 
    +
    e^{- iq\cdot x}\,a_-(\vec q\,)
  \right]\quad
\end{eqnarray}
with $q_0$ the energy of the meson and $a^\dagger$ and $a$ are the usual creation and
annihilation operators.
Here $e$ is the positive elementary charge, $m_\pi$ is the $\pi^+$-meson mass, 
$f$ is the $\pi$-N coupling constant, $\vec \sigma$ denotes the Pauli spin matrix vector and
$\tau_-$ the isospin operator.  Dressler {\it et al.} \cite{Dress} show that the KR
term gives nearly the entire cross section for pion energies in the range
considered here.  Due to
cancellations other terms in the production operator i.e. terms required by 
gauge invariance, do not contribute significantly at low pion energies.  
With $E_{\rm cm}$ as the
incident c.m.~energy
the response function $\cal R$($E_{\rm cm}$) for this process
is
\begin{eqnarray}
  {\cal R}({E_{\rm cm}})
  =
  \frac{1}{6}\,
  \sum_{M_d,\lambda}\sum_f 
  \left|
    \langle\,f\,|\,
    \tilde{\cal O}(\vec k,\lambda)\,
    |\,D,M_d\,\rangle \right|^2\, 
  \delta({E_{\rm cm} -E_f})\quad,\label{eqn:3}
\end{eqnarray}
where $|\,D,M_d\,\rangle$ denotes the deuteron ground state with polarization $M_d$ and
\begin{eqnarray}
  \tilde{\cal O}(\vec k,\lambda)
  =
  \int\!{\rm d}^3q\,\,
  {\cal O}(\vec k,\lambda,\vec q\,)\,a_+^\dag(\vec q\,)
\end{eqnarray}
with
\begin{eqnarray}
  {\cal O}(\vec k,\lambda,\vec q\,)
  =
  -i\,e\,
  \left(\frac{f}{m_\pi}\right)\,
  \frac{1}{\sqrt{8 \pi^3}}\,
  \frac{1}{\sqrt{2 q_0}}\,
  \sum_j\,\hat\epsilon_\lambda\cdot
  \vec\sigma(j)\,\tau_-(j)\,e^{i\, \vec x_j\cdot(\vec k - \vec q\,)}
  \quad.
\end{eqnarray}
In  Eq.~(\ref{eqn:3}) above  
$|\,f\,\rangle$ denotes the 
wave function of the relative motion 
of the $nn\pi^+$ system with  energy $E_f$, and $\sum_f$
indicates an integration over all relative momenta and a sum over all final nucleon spins.
The inclusive cross section is related to ${\cal R}(E_{\rm cm})$ by
\begin{eqnarray}
  \sigma(E_{\rm cm})
  =
  \frac{2\pi^2}{k}{\cal R}(E_{\rm cm})
  \quad.
\end{eqnarray}
Relative momenta in the final state are taken as 
\begin{eqnarray}
  \vec p_x
  &=&
  (\vec p_1 - \vec p_2)/2
  \\
  \vec p_y
  &=&
  -\frac{m_\pi}{M}\,(\vec p_1 + \vec p_2) + \frac{2m_n}{M}\,\vec p_\pi 
\end{eqnarray}
where 
$\vec p_1,\vec p_2,$ and $\vec p_\pi$ 
are the momenta of the
final state neutrons and pion, respectively, and $M$ is the total mass $2m_n + m_\pi$
with the mass of the neutron $m_n$.
As implied by this separation we are restricting ourselves to non-relativistic
kinematics for both the nucleons and the pion.  This allows the separation of
the hamiltonian into Jacobi coordinates for the NN$\pi$ three body system.
In terms of these quantities the energy conserving
$\delta$-function appearing in ${\cal R}(E_{\rm cm})$ takes the detailed form
\begin{eqnarray}
  \delta(E_{\rm cm}-E_f)
  =
  \delta\left(
  E_{\rm cm} -(2 m_n + m_\pi ) - \frac{p_x^2}{m_n} - \frac{p_y^2}{{2 \mu}}\right)\quad,
\end{eqnarray}
where $\mu$ is the reduced 
two-neutron-pion mass.
Finally it is more convenient
to use the c.m.~energy above threshold 
\begin{eqnarray}
  {\cal W} 
  = 
  k + \frac{k^2}{2 m_d} - (2 m_n + m_\pi - m_d ) \geq 0
  \nonumber
\end{eqnarray}
where $m_d$ is the mass of the deuteron.
The LIT, referred to hereafter as the transform, of the response
function ${\cal R}({\cal W})$ is then defined with $\sigma_R,\,\sigma_I>0$ as
\begin{eqnarray}
  L(\sigma_R,\sigma_I)
  &=&
  \int_0^\infty\!{\rm d}{\cal W}\,
  \frac{{\cal R}({\cal W})}{({\cal W}-\sigma_R)^2+\sigma_I^2}
  \\
  &=&
  \frac{1}{6}\sum_{M_d,\lambda}\,\int_0^\infty\!{\rm d}{\cal W}\,
  \langle\, D,M_d\,|\,
  \tilde{\cal O}^\dag(\vec k,\lambda)\,\frac{\delta({\cal W} - H )}
  {(H - \sigma_R)^2 + \sigma_I^2}\,\tilde{\cal O}(\vec k,\lambda)\,
  |\,D,M_d\,\rangle .
 \end{eqnarray} 
Because of the relation of $k$ to ${\cal W}$, a straightforward integration
of the above equation would leave the operator $\tilde{\cal O}(\vec k,\lambda)$
depending in a complicated way on the Hamiltonian $H$. 
Therefore, we proceed by setting $\vec k$ appearing in the operator $\tilde{\cal O}$ 
to a constant arbitrarily chosen "pseudo-momentum" $\vec k_{\rm p}$. 
As a result we introduce a new transform $L_{k_{\rm p}}$, which depends on $k_{\rm p}$ and takes the form
\begin{eqnarray}
  L_{k_{\rm p}}(\sigma_R,\sigma_I)
  &=&
  \frac{1}{6}\sum_{M_d,\lambda}\,
  \langle\, D,M_d\,|\,
  \tilde{\cal O}^\dag(\vec k_{\rm p},\lambda)\,
  \frac{1}{(H - \sigma_R)^2 + \sigma_I^2}\,\tilde{\cal O}(\vec k_{\rm p},\lambda)\,
  |\,D,M_d\,\rangle
  \\
  &=&
  \frac{1}{6}\sum_{M_d,\lambda}\,
  \langle\,{\tilde\psi}_{M_d,\lambda}(\sigma_R,\sigma_I,k_{\rm p})\,|\,
  {\tilde\psi}_{M_d,\lambda}(\sigma_R,\sigma_I,k_{\rm p})\,\rangle\quad,
\end{eqnarray}
where
the Lorentz function $\tilde\psi$ is solution of the inhomogeneous equation
\begin{eqnarray}
  (H - \sigma_R + i\sigma_I)\,|\,{\tilde\psi}_{M_d,\lambda}(\sigma_R,\sigma_I,k_{\rm p})\,\rangle
  =
  {\tilde{\cal O}}(\vec k_{\rm p},\lambda)\,|\,D,M_d\,\rangle
\end{eqnarray}
with
\begin{eqnarray}
  \langle\,\vec p_x,\vec p_y\,|\,H\,|\,{\vec p}_x^{\ \prime},\vec p_y^{\ \prime}\,\rangle
  =
  \left[
    \left(
      \frac{p_x^2}{m_n}
      +
      \frac{p_y^2}{2\mu}
      \right)\,\delta (\vec p_x - \vec p_x^{\ \prime})
      + 
      V(\vec p_x,\vec p_x^{\ \prime})
      \right]\,
      \delta (\vec p_y - \vec p_y^{\ \prime})\quad.
\end{eqnarray}

The inverse to the transform $L_{k_{\rm p}}(\sigma_R,\sigma_I)$ will lead to
a corresponding response function ${\cal R}_{k_{\rm p}}({\cal W})$ and 
in turn will yield the correct
cross section for $k_{\rm p}=k$.  
Consequently for each photon energy the calculation of the transform
and its inversion must be repeated. 

As mentioned earlier the simple model here only includes an NN potential, 
i.e.,~$V(\vec p_x,\vec p_x^{\ \prime}$).  The kinetic energy of the meson with respect
to the $nn$ pair appears in $H$ as the term proportional to 
$p_y^2$. 
Meson rescattering could be included by adding to $H$ appropriate potentials.
Although that would considerably complicate the numerical aspect, it would
not affect the question posed here, namely: can the LIT be
inverted when the response function arises from a particle production
process? 
A problem is the slow fall-off of the response function for
high ${\cal W}$ values.
From the definition of the transform it is clear that
the response must behave asymptotically like ${\cal W}^{1-x}$ where $x>0$ in order for the
integral to converge. 
In the case of nuclear photoabsorbtion without pion production 
the response function falls off very rapidly for 
increasing ${\cal W}$ resulting in $L(\sigma_R,\sigma_I)$ also falling rapidly for large
$\sigma_R$. 
Inversion then gives an accurate account of ${\cal R}({\cal W})$ over
its entire range by calculating
$L(\sigma_R,\sigma_I)$ for a finite number of $\sigma_R$ values.  That 
the response function for inclusive photoproduction may have a significantly
different asymptotic behavior can be seen from the basic Kroll-Rudermann \cite{KR}
cross-section for $p(\gamma,\pi^+)n$ 
\begin{eqnarray}
  \sigma
  =
  2\,\alpha\, 
  \left(\frac{f}{m_\pi}\right)\,
  \left(\frac{q}{k}\right)\,
  \frac{E_n(q)\,E_p(k)}{E_{\rm cm}^2}
\end{eqnarray}
where $E_N$ are the nucleon energies, $E_{\rm cm}$ is the c.m.~energy $k + E_p(k)$ and
$\alpha$ the fine structure constant.  
This cross section approaches a constant
for large $E_{\rm cm}$ and the response ${\cal R}(E_{\rm cm})$ rises linearly in $E_{\rm cm}$. 
In the next section it will be seen that
the pion production response function for a finite nucleus is tempered at large $E_{\rm cm}$ by
structure effects. 
Nevertheless it still rises over a large energy region
thereby requiring a different approach for the  inversion 
of the transform.

\section{Results}

Details of the model and the parameters used are given in the Appendix.  There
it is seen that the vector $|\,{\tilde\psi}_{M_d,\lambda}(\sigma_R,\sigma_I,k_{\rm p})\,\rangle$
can also be labeled by the final neutron-neutron spin and that the response
function is therefore a sum of singlet and triplet contributions. 
The separable potential used here only has scattering in the spin $S=0$, isospin $T=1$ channel
while the $S=1$, $T=1$ final state consisting of odd partial waves
is non-interacting.  Fig.~1 shows the 
transform $L_{k_{\rm p}}(\sigma_R,\sigma_I)$ for the case $\sigma_I=k_{\rm p}=10$ MeV.  
One notes that $L_{k_{\rm p}}$ reaches a maximum at 
$\sigma_R\approx 200$ MeV and then falls very slowly with increasing
$\sigma_R$.  Also shown in this figure is the response function 
${\cal R}_{k_{\rm p}}({\cal W})$ for $k_{\rm p}=10\,\mbox{MeV}$.
The response function behaves similarly to $L$ as one would expect
since $L(\sigma_R,\sigma_I)$ samples ${\cal R}$ mainly from a region centered
at ${\cal W}=\sigma_R$. We note that ${\cal R}_{k_{\rm p}}({\cal W})$ does not 
rise linearly with energy as in the case of $p(\gamma,\pi^+)n$ but falls off
slowly with energy after reaching a maximum. Unfortunately, the fall-off of
$L$ with $\sigma_R$ is
so slow and covers such a large energy range that inverting it to
obtain ${\cal R}({\cal W})$ over the full range, as was possible in the earlier
photoabsorption calculations, is not only very difficult but
largely physically meaningless because our model is only valid 
for non-relativistic mesons. 
In fact the present model is only sensible for energies ${\cal W}$ not exceeding
approximately $40$ MeV 
which corresponds to the maximal pion  energy 
still being non-relativistic. Our aim then is to
calculate $L$ only for the segment $0\leq\sigma_R\leq 40$ MeV and then invert it
to extract ${\cal R}_{k_{\rm p}}({\cal W})$ for ${\cal W}$ in the same energy range. 
Since ${\cal R}$
can be calculated directly the error can be easily assessed.

The inversion process we use has already been described in \cite{elo6} but
a brief account is as follows.  The response function ${\cal R}_{k_{\rm p}}({\cal W})$ is written as a sum  
\begin{eqnarray}
  {\cal R}_{k_{\rm p}}({\cal W})
  =
  \sum_{i=1}^N\,\alpha_i\,{\cal W}^{S+2}\,e^{-\beta_i {\cal W}}
  \equiv
  \sum_{i=1}^N\,{\cal R}_{k_{\rm p}}^i({\cal W})
\end{eqnarray}
and the parameters $\alpha_i$ and $\beta_i$ are determined by fitting
\begin{eqnarray}
  L_{k_{\rm p}}^{\rm fit}(\sigma_R,\sigma_I)
  =
  \sum_{i=1}^N\,\int_0^\infty\!{\rm d}{\cal W}\,
  \frac{{\cal R}_{k_{\rm p}}^i({\cal W})}
  {({\cal W}-\sigma_R)^2 + \sigma_I^2}
\end{eqnarray}
to $L$ as computed from Eqs.~(\ref{eqn:app1})-(\ref{eqn:app2}) 
of 
the Appendix. Note that the threshold behavior
${\cal W}^{S+2}$, where $S$ is the total spin, is appropriate to the
photoproduction process \cite{Gupta} and differs slightly from the form used
in the earlier photoabsorption calculations. It turns out however that
the quality of fit is nearly independent of which threshold behaviour
is used. 
Fig.~2 shows the quality of the
fit obtained using $N=10$.  
With the parameters so determined one can 
calculate the response functions and finally the inclusive cross section.
As mentioned earlier this model problem is simple enough that 
the response functions are easily calculated in the traditional
manner \cite{Dress,Gupta}.
Fig.~3 shows the relative error between
the response function as calculated from the LIT technique 
( ${\cal R}_{\rm fit}$ ) and 
the one  calculated in the traditional manner.  
Over most
of the energy range between $0\le{\cal W}\le 40$ MeV the relative error is
in the $1\%$ range.  
However near the threshold, i.e.~${\cal W}<2$ MeV the response
function tends to zero for ${\cal W}\longrightarrow 0$ and limited 
numerical accuracy produces an
exaggerated relative error in this region.  
These effects are too small to be
visible in a plot of the cross section however. Finally our computed 
cross section along with data of Booth {\it et al.} \cite{Booth} is shown
in Fig.~4.  The results of using the LIT technique
are indistinguishable from the traditional methods in \cite{Dress,Gupta}.

It is instructive to see the differences in the response functions that
occur if one uses transforms calculated in  different ranges of 
$\sigma_R$.  One expects that because of the nature of the Lorentz transform
that one should only have to calculate the transform up to $\sigma_R=40$\,MeV, 
as was done above,
if the response function was only required in the energy range
$0\le{\cal W}\le 40$ MeV.  To show this we calculate the transform in four
ranges $0\le\sigma_R\le\sigma_{R_{\rm max}}$ where $\sigma_{R_{\rm max}}=20,40,60$, and
$70$\,MeV. The respective response functions obtained by inverting each of these
transforms is denoted by ${\cal R}_{k_{\rm p}}({\cal W}|\sigma_{R_{\rm max}})$.
Fig.~5 shows the error in the $\sigma_{R_{\rm max}}$= 20,40,60 MeV cases relative
to the $\sigma_{R_{\rm max}}$= 70 MeV case.  One notes that the $\sigma_{R_{\rm max}}$= 20
MeV case only is accurate for energies up to 20 MeV, that the 
$\sigma_{R_{\rm max}}$= 40 MeV case has a less than 1$\%$ error at
${\cal W}$= 40 MeV,
and that the $\sigma_{R_{\rm max}}$= 60 MeV case has only a very small error
up to 60 MeV.  

\section{Conclusions}

We have shown that the LIT technique can be used to
calculate inclusive meson photoproduction cross sections in the non-relativistic
regime. Rather than fitting the transform over its entire range as was possible
in earlier photoabsorbtion calculations one fits here only the low energy
segment to obtain the low energy response functions.  The model calculation
used here shows that the response functions thus obtained are as accurate
as numerical techniques will allow. 
Our next step will be to add a pion-nucleon interaction to the Hamiltonian
in order to take account of pion scattering effects.  The dynamical model
of Darwish, Arenhoevel, and Schwamb \cite{DAS} would provide a benchmark
calculation against which to further check our method.
It would also be of interest to apply these types of calculations
to higher $A$ nuclei such as $^3$He, $^3$H, or $^4$He.  
The LIT technique 
extends readily to these cases as well as 
being able to handle realistic potentials including Coulomb effects.
One should expect to be able to study meson-rescattering
effects with realistic nuclear models in a variety of light nuclei.

\section*{Acknowledgment}

We would like to thank H.~Arenh{\"o}vel for valuable comments.
All authors acknowledge support from the Italian Ministery of 
Research (MURST). In addition the work of C.R.~and E.L.T. is supported 
by the National Science and Engineering Research Council of Canada.
\newpage

\section*{Appendix}
\begin{appendix}
\setcounter{equation}{0}

For the NN interaction we use the separable model of
Y.~Yamaguchi \cite{Yam}, which is a pure $S$-wave interaction
\begin{eqnarray} 
  V(\vec{p},\vec{p}\,')
  & = &
  -\lambda_{0}\,g_{0}(\vec{p}\,)\,g_{0}(\vec{p}\,') 
  \,\frac{1}{4}\left(1-\vec{\sigma}_1\cdot\vec{\sigma}_2\right)\nonumber\\
  && 
  - \lambda_{1}\,g_{1}(\vec{p}\,)\,g_{1}(\vec{p}\,')
  \,\frac{1}{4}\left(3+\vec{\sigma}_1\cdot\vec{\sigma}_2\right)
  \quad,
  \\\nonumber\\
  g_{S}(\vec{p}\,)
  & = &
  \frac{1}{\vec{p}\,^2+\beta_S^2}
  \quad S\in\{0,1\}\,.
\end{eqnarray}
Here the labels $0$ and $1$ refer to the spin-singlet
and spin-triplet parts respectively. 
The following more up to date constants for this model have been 
taken from \cite{Dress}:
\begin{eqnarray}
  \alpha  
  &=&  0.2316\,{\rm fm}^{-1}
  \quad,\quad
  \beta_0  
  =  1.129\,{\rm fm}^{-1} 
  \quad,\quad
  \beta_1  
  =  
  1.392\,{\rm fm}^{-1}
  \quad,\\
  \lambda_0 
  & = & 
  0.02774\,{\rm fm}^{-2}
  \quad,\quad
  \lambda_1
  = 
  \frac{\beta_1\,(\alpha+\beta_1)^2}{m_n\,\pi^2}
  \quad.
\end{eqnarray}
These
constants fit the deuteron binding energy, the experimental 
values for the singlet and triplet scattering lengths 
and the singlet effective range.

Using this separable NN interaction one obtains for the
deuteron ground state
\begin{eqnarray}
  \psi_d(\vec{p}\,)
  =
  \frac{\sqrt{\alpha\,\beta_1\,(\alpha+\beta_1)^3}}{\pi}
  \,\frac{1}
  {\left(\vec{p}\,^2+\alpha^2\right)\,
    \left(\vec{p}\,^2+\beta_1^2\right)}\,
\end{eqnarray}
and a binding energy of $2.224$ MeV.

The Lorentz function $\tilde{\psi}_{M_d,\,\lambda}(\sigma,\,k_{\rm p},\,\vec{p}_x,\,\vec{p}_y)$
can be decomposed into its spin components as
\begin{eqnarray}
  \tilde{\psi}_{M_d,\,\lambda}(\sigma,\,k_{\rm p},\,\vec{p}_x,\,\vec{p}_y)
  =
  \sum_{S=0,1}
  \tilde{\psi}_{M_S}^{S}(\sigma,\,k_{\rm p},\,\vec{p}_x,\,\vec{p}_y)
\end{eqnarray}
where $M_S=M_d+\lambda$ with $M_d$ and $\lambda$ the deuteron and photon polarization,
respectively,  and the function 
$\tilde{\psi}_{M_S}^{S}(\sigma,\,k_{\rm p},\,\vec{p}_x,\,\vec{p}_y)$ is a solution of
\begin{eqnarray}
  \left[
    \frac{p_x^2}{m_n}+ \frac{p_y^2}{2\mu}-\sigma
  \right]\,
  \tilde{\psi}^{S}_{M_S}(\sigma,\,k_{\rm p},\,\vec{p}_x,\,\vec{p}_y)
  -  
  \lambda_{S}\,g_{S}(p_x)\,
  C^{S}_{M_S}(\vec{p}_y,\,\vec{\Delta})
  &&
  \nonumber\\\nonumber\\
  \qquad\qquad  =
  \Xi^{S}_{M_S,M_d,\lambda}
  \,
  {\cal F}^{S}_{\vec{\Delta}}(\vec{p}_x)
  \quad,\quad S\in\left\{0,1\right\}\quad,
 \end{eqnarray} 
where 
\begin{eqnarray}
  C^{S}_{M_S}(\vec{p}_y,\,\vec{\Delta})
  & = &
  \int\!{\rm d}^3p\,g_{S}(p)\,
  \tilde{\psi}^{S}_{M_S}(\sigma,\,k_{\rm p},\,\vec{p},\,\vec{p}_y)\quad,
  \\
  {\cal F}^{S}_{\vec{\Delta}}(\vec{p}\,)
  & = &
  \psi_d(\vec{p}-\vec{\Delta})
  +
  (-1)^{S}
  \psi_d(\vec{p}+\vec{\Delta})
  \quad,\\
  \Xi^{S}_{M_S,M_d,\lambda} 
  & = &
   -i\,\frac{3\,\sqrt{2}\,e\,f}{m_\pi}
  \,
  (-1)^{1+S-M_S}\,
  \sqrt{2S+1}\,
  \begin{pmatrix}
    S    & 1   & 1 \\ 
    -M_S & \lambda & M_d 
  \end{pmatrix}
  \left\{
    \begin{matrix}
      \frac{1}{2} & S          & \frac{1}{2} \\
      1           & \frac{1}{2} & 1
    \end{matrix}
  \right\}\quad.
 \end{eqnarray} 
We take $f^2/4\pi$=0.078.

The constant $C^{S}_{M_S}(\vec{p}_y,\,\vec{\Delta})$ with respect to $p_x$ is
\begin{eqnarray}
  C^{S}_{M_S}(\vec{p}_y,\,\vec{\Delta})
  & = &
  {m_n\,\Xi^{S}_{M_S,M_d,\lambda}\,
    \int\!{\rm d}^3p\,
    \frac{g_{S}(p)\,{\cal F}^{S}_{\vec{\Delta}}(\vec{p}\,)}
    {p^2+\gamma^2-m_n\,\sigma}}
  \left[
    {
      1
      -
      \frac{m_n\,\pi^2\,\lambda_{S}}
      {\beta_{S}\,
        \left(\beta_{S}+\sqrt{\gamma^2-m_n\,\sigma}\right)^2}
      }
    \right]^{-1}
 \end{eqnarray} 
with 
$
  (m_n /\mu)/2\,\vec{p\,}_y^2
      \equiv \gamma^2(\vec{p}_y)
$.
In the case of $S=1$ this constant vanishes.

The solution of the Lorentz-equation therefore is
\begin{eqnarray}\label{eqn:app1}
  \tilde{\psi}^{S}_{M_S}(\sigma,\,k_{\rm p},\,\vec{p}_x,\,\vec{p}_y)
  = 
  \frac{m_n\,\Xi^{S}_{M_S,M_d,\lambda}\,
    {\cal F}^{S}_{\vec{\Delta}}(\vec{p}_x)
    +
    m_n\,\lambda_{S}\,g_{S}(p_x)\,C^{S}_{M_S}(\vec{p}_y,\,\vec{\Delta})
    }
  {p_x^2+\gamma^2-m_n\,\sigma}\quad
 \end{eqnarray} 

and the transform $L_{k_{\rm p}}(\sigma_R,\sigma_I)$
is
\begin{eqnarray}\label{eqn:app2}
  L_{k_{\rm p}}(\sigma_R,\sigma_I) 
  = 
  \frac{1}{6}  
  \sum_{S,\,M_d,\,\lambda} \langle\tilde{\psi}^{S}_{M_{S}}\,|\,\tilde{\psi}^{S}_{M_{S}}\rangle
  \quad.
\end{eqnarray}

\end{appendix}

\newpage

\section{Bibliograhy}

\newpage

\begin{figure}
  \centerline{\epsfig{file=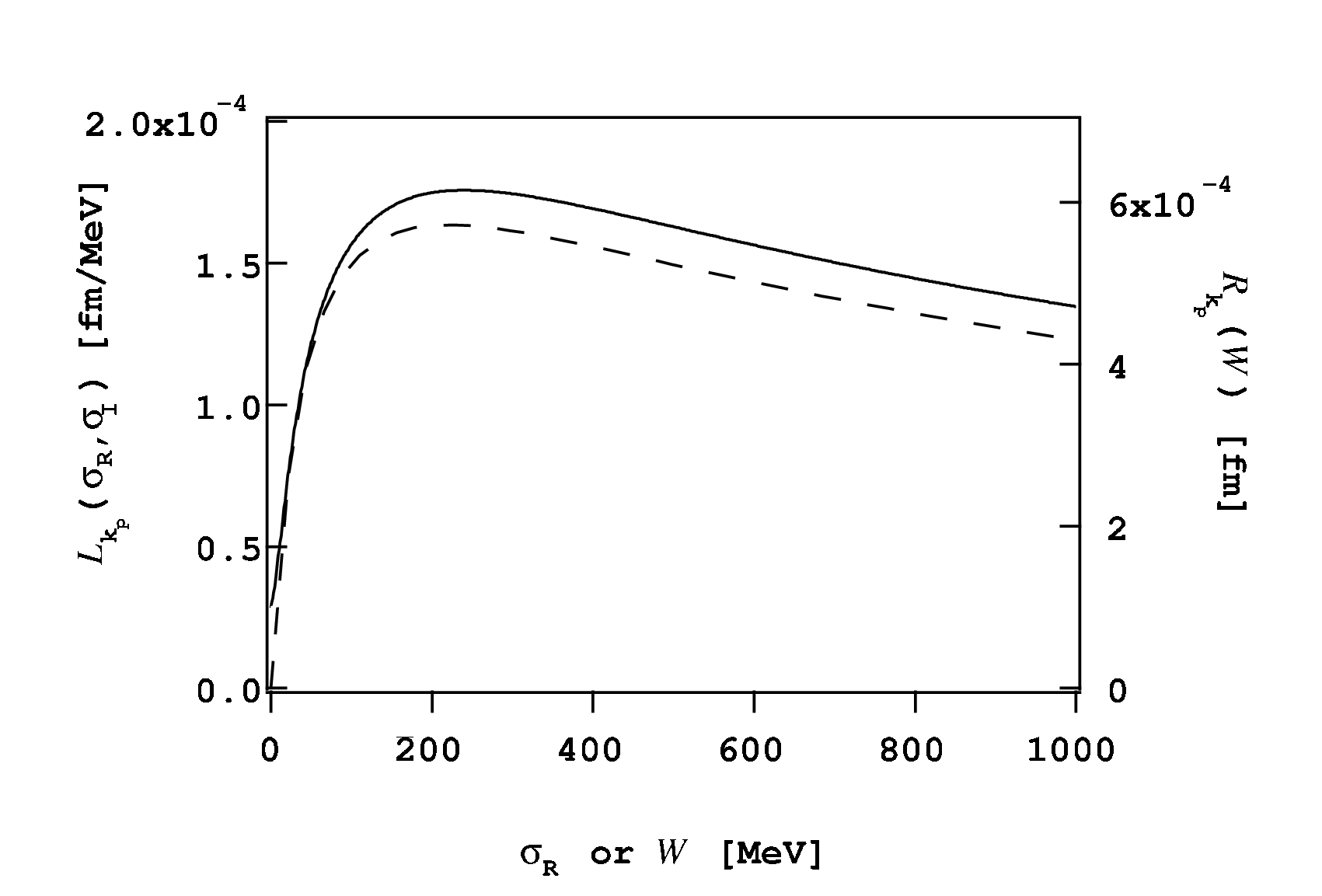,width=\textwidth,angle=0}}
  \caption{ $L_{k_{\rm p}}(\sigma_R,\sigma_I)$ (solid line) and
 ${\cal R}_{k_{\rm p}}({\cal W})$ (dashed line) shown for
    $\sigma_I=k_{\rm p}= 10$ MeV.  }
\end{figure}

\begin{figure}
  \centerline{\epsfig{file=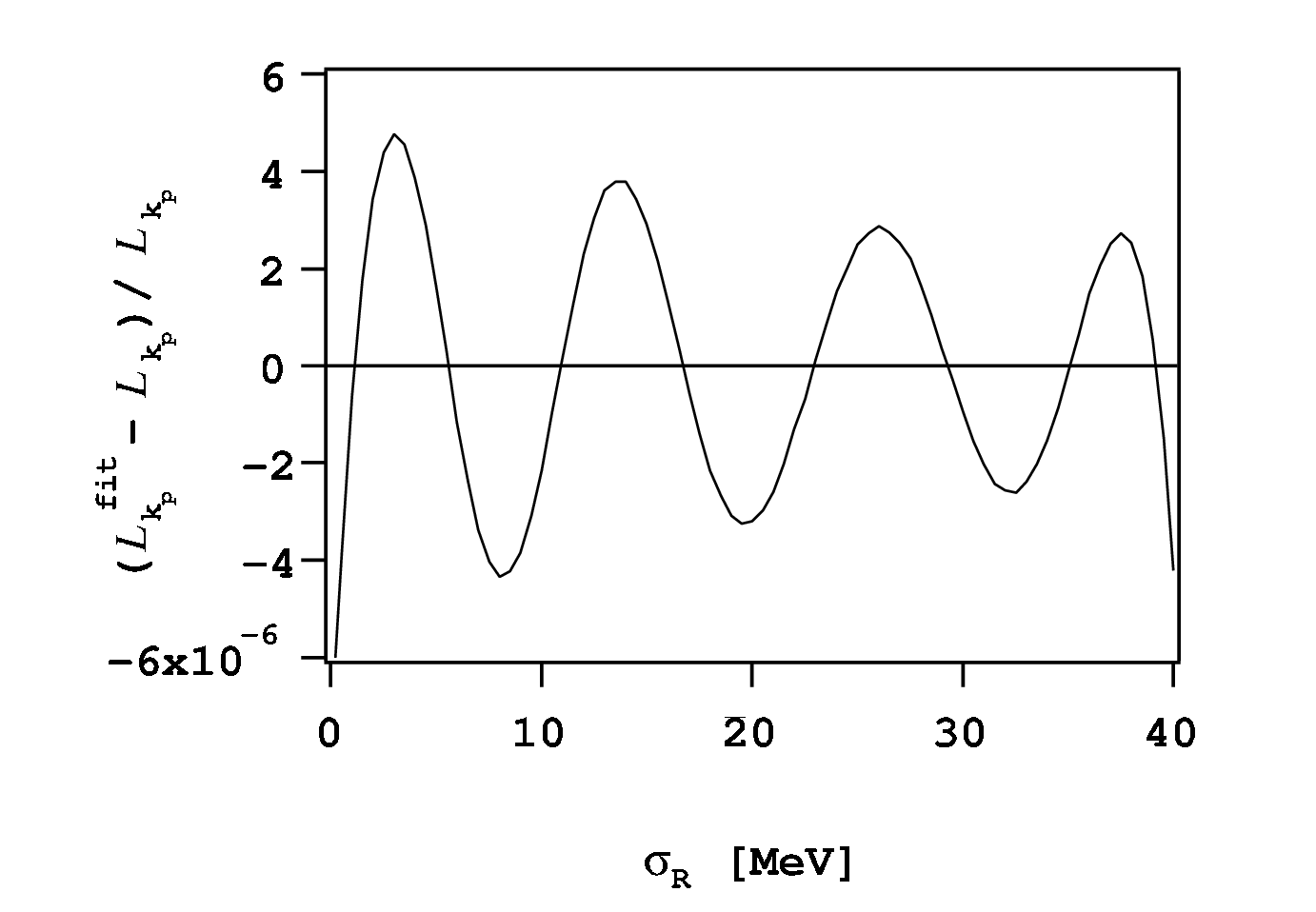,width=\textwidth,angle=0}}
  \caption{ Relative error of the fit $L^{\rm fit}_{k_{\rm p}}(\sigma_R,\sigma_I)$ compared
    to the computed transform $L_{k_{\rm p}}(\sigma_R,\sigma_I)$.  Here we have used
    $\sigma_I=k_{\rm p} = 10$ MeV for illustration.
}
\end{figure}

\begin{figure}
  \centerline{\epsfig{file=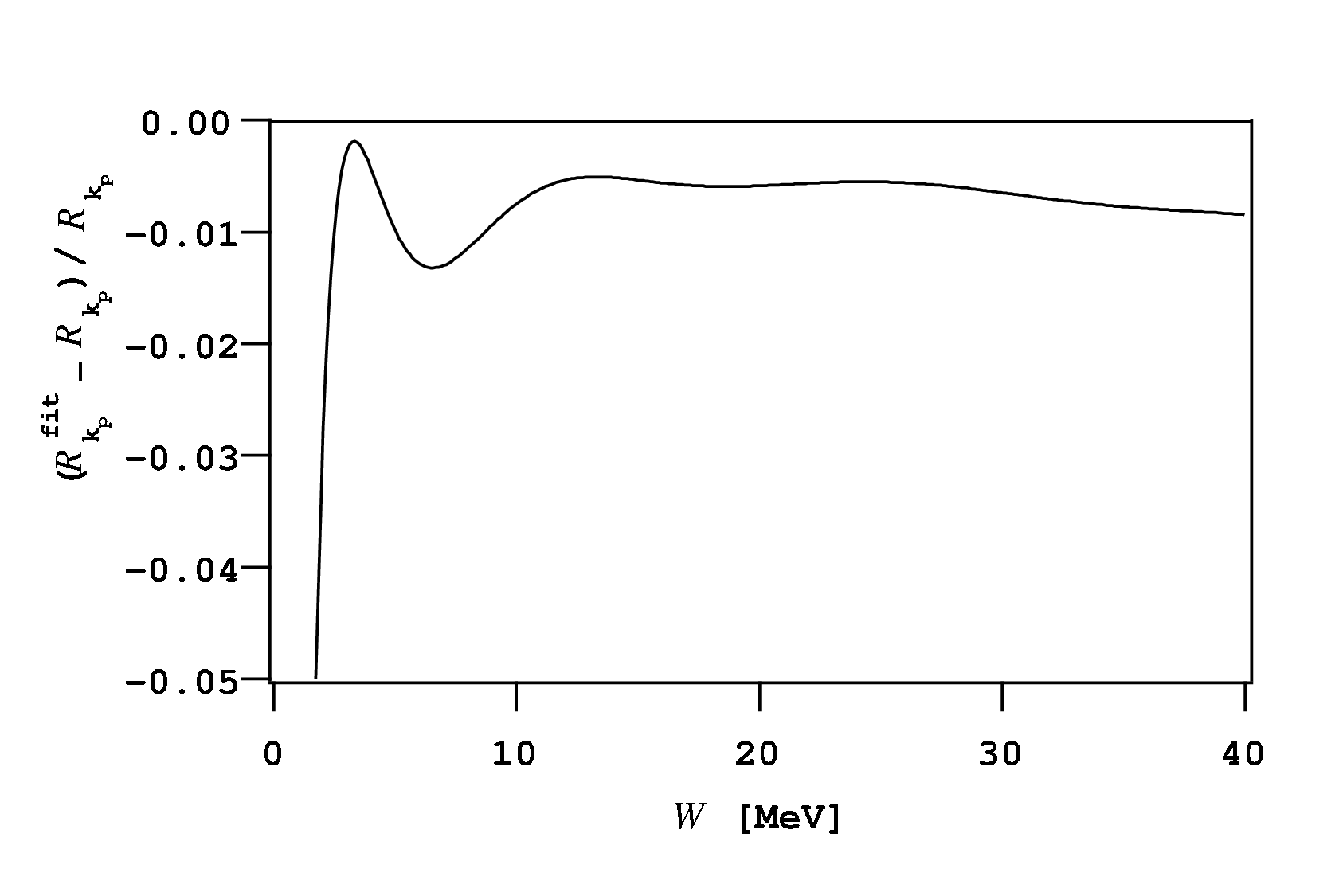,width=\textwidth,angle=0}}
  \caption{Relative error between the response function ${\cal R}_{k_{\rm p}}^{\rm fit}({\cal W})$ 
    and ${\cal R}_{k_{\rm p}}({\cal W})$ computed in the traditional method for $\sigma_I=k_{\rm p} = 10$ MeV.}
\end{figure}

\begin{figure}
\centerline{\epsfig{file=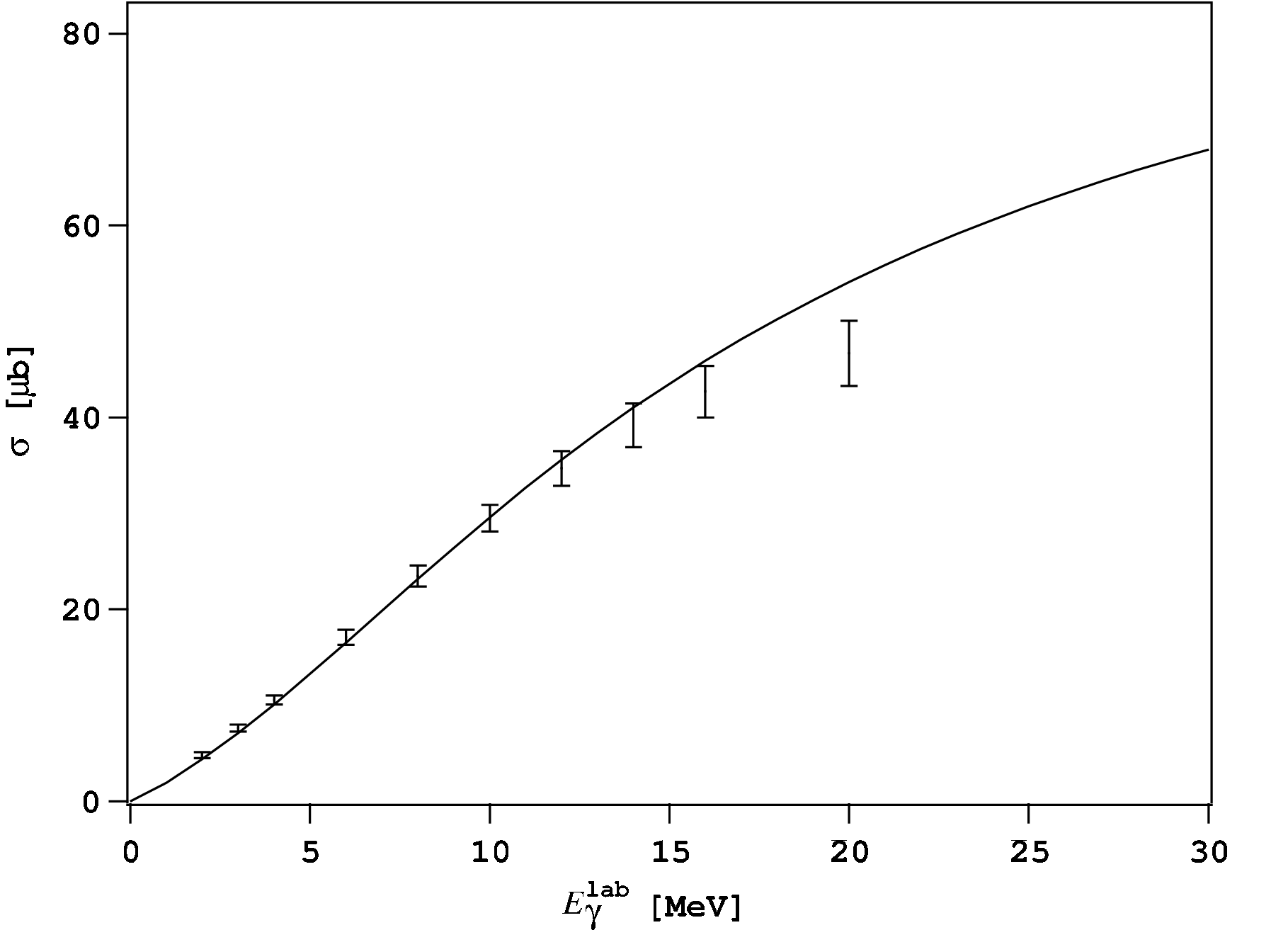,width=\textwidth,angle=0}}
\caption{ Calculated total cross section for D$(\gamma,\pi^+)nn$ shown together
  with the data of \cite{Booth}.  The cross sections calculated either by
  the LIT technique or the traditional method \cite{Dress,Gupta} are
  indistinguishable. }
\end{figure}

\begin{figure}
\centerline{\epsfig{file=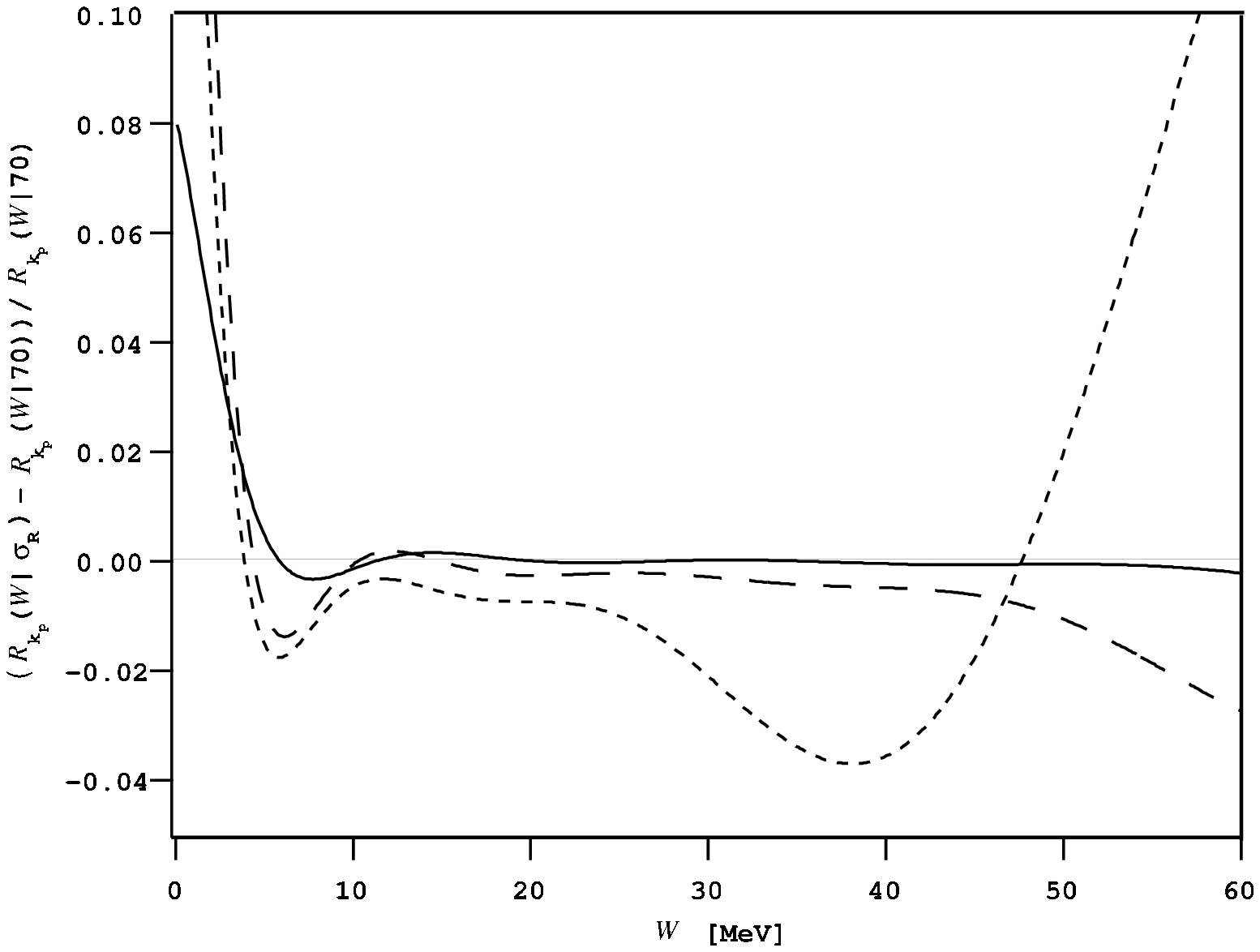,width=\textwidth,angle=0}}
\caption{Relative errors in response functions for $\sigma_I=k_{\rm p} = 10$ MeV with respect to the 
$\sigma_{R_{\rm max}}$=70 MeV case: $\sigma_{R_{\rm max}}$= 60 MeV (solid line),
$\sigma_{R_{\rm max}}$= 40 MeV (dashed line), $\sigma_{R_{\rm max}}$= 20 MeV (dotted line).}
\end{figure}

\end{document}